\documentclass[twocolumn,prl,floatfix]{revtex4}
\usepackage{graphicx}
\usepackage{amssymb,amsmath,latexsym}
\usepackage{bm}
\usepackage[dvips]{epsfig}

\usepackage{color}

\newcommand{\vhat}[1]{\hat{\bm{\mathrm{#1}}}}

\begin{document}

\title{Chirality from Interfacial Spin-Orbit Coupling Effects in Magnetic Bilayers}
\author{Kyoung-Whan Kim$^{1,2}$}
\author{Hyun-Woo Lee$^2$}\email{hwl@postech.ac.kr}
\author{Kyung-Jin Lee$^{3,4}$}
\author{M. D. Stiles$^5$}
\affiliation{ $^1$Basic Science Research Institute, Pohang
University of Science and Technology, Pohang, 790-784, Korea
\\
 $^2$Department of Physics, Pohang University of
Science and Technology, Pohang, 790-784, Korea
\\
$^3$Department of Materials Science and Engineering, Korea
University, Seoul, 136-701, Korea
\\
$^4$KU-KIST Graduate School of Converging Science and Technology,
Korea University, Seoul 136-713, Korea
\\
$^5$Center for Nanoscale Science and Technology, National
Institute of Standards and Technology, Gaithersburg, Maryland
20899-6202, USA }

\date{\today}
%

\begin{abstract}
%
As nanomagnetic devices scale to smaller
sizes, spin-orbit coupling due to the broken structural inversion
symmetry at interfaces becomes increasingly important.
Here we study interfacial spin-orbit coupling effects in
magnetic bilayers using a simple Rashba model.
The spin-orbit coupling introduces chirality into the
behavior of the electrons and through them into the energetics of the
magnetization.  In the derived form of the magnetization dynamics,
all of the contributions that are linear
in the spin-orbit coupling follow from
this chirality, considerably
simplifying the analysis.
For these systems, an important consequence is a correlation between the Dzyaloshinskii-Moriya
interaction and the spin-orbit torque. We use this correlation to
analyze recent experiments.
\end{abstract}


\maketitle

Magnetic bilayers that consist of an atomically thin ferromagnetic layer
(such as Co) in contact with a nonmagnetic layer (such as Pt) with strong
spin-orbit coupling have emerged as prototypical systems that exhibit very
strong spin-orbit coupling effects.
Strong spin-orbit coupling can enhance the efficiency of the electrical control of
magnetization. A series of recent
experiments~\cite{Miron:2010,Miron:2011a,Miron:2011b,Liu:2012a} on
magnetic bilayers report dramatic effects such as anomalously fast
current-driven magnetic
domain wall motion~\cite{Miron:2011a} and reversible
switching of single ferromagnetic layers by in-plane
currents~\cite{Miron:2011b,Liu:2012a}.
Strong spin-orbit coupling can introduce chirality into the
magnetic ground state~\cite{Bode:2007,Heinze:2011}.
This chirality is predicted~\cite{Thiaville:2012}
to boost the electrical control of magnetic degrees of
freedom even further as has been confirmed in two
experiments~\cite{Emori:2013,Ryu:2013}.

Interfaces lack structural
inversion symmetry, allowing interfacial spin-orbit coupling to
play an expanded role.  In magnetic bilayers, it generates
various effects including the Dzyaloshinskii-Moriya (DM)
interaction~\cite{Dzyaloshinskii:1957,Moriya:1960,Fert:1990} and
the spin-orbit torque~\cite{Obata:2008,Matos-Abiague:2009,Wang:2012,Kim:2012a,Pesin:2012,vanderBijl:2012}.
Here, we examine a simple Rashba model of the interface region. We
compute the equation of motion for a magnetization texture
$\hat{\bf m}({\bf r})$ by integrating out the electron degrees of
freedom.
%
We report two main findings.
The first is the correlation between the DM
interaction and the spin-orbit torques.
Spin-orbit torques arise from interfacial spin-orbit coupling but
also from the bulk spin Hall effect, and the importance of each
contribution is hotly
debated~\cite{Miron:2011b,Liu:2012b,Kim:2013,Haazen:2013,Freimuth:2013}.
The correlation we find opens a way to quantify the contribution
from interfacial spin-orbit coupling by measuring the DM
interaction, allowing one to disentangle the two contributions.

The second finding is that all linear effects of the interfacial spin-orbit
coupling, including the DM interaction and
the spin-orbit torque, can be
captured through a simple mathematical construct, which we call a
chiral derivative.
The chiral derivative also shows in the equation of motion how
each contribution that is linear in the spin-orbit coupling
corresponds to a contribution that is present even in the absence
of spin-orbit coupling.
This
correspondence provides a simple way to quantitatively predict and
understand a wide variety of interfacial spin-orbit coupling
effects allowed by symmetry~\cite{vanderBijl:2012}.
%
%
In the last part of the Letter, we discuss briefly the extension to realistic
situations, which go beyond the simple Rashba model.

Our analysis begins with the two-dimensional (2D) Rashba Hamiltonian
\begin{eqnarray}
{\cal H}&=&{\cal H}_{\rm kin}+{\cal H}_{\rm R}+{\cal H}_{\rm exc}+{\cal H}_{\rm imp}
\nonumber\\
&=&\frac{{\bf p}^2}{2m_e} + \frac{\alpha_{\rm
R}}{\hbar}\bm{\sigma}\cdot({\bf p}\times \hat{\bf
  z})
+J\bm{\sigma}\cdot \hat{\bf m}+{\cal H}_{\rm imp},
\label{eq:RSOC-Hamiltonian}
\end{eqnarray}
where ${\bf p}$ is the 2D electron momentum in the $xy$ plane, the vector
$\bm{\sigma}$ of the Pauli matrices  represents the electron spin, and
$|\hat{\bf m}({\bf r})|=1$.
${\cal H}$ is a minimal
model~\cite{Obata:2008,Matos-Abiague:2009,Wang:2012,Kim:2012a,Pesin:2012,vanderBijl:2012}
for electronic properties of the interface region between the ferromagnetic
and nonmagnetic layers in magnetic bilayers, and captures the broken
symmetries; ${\cal H}_{\rm exc}$ breaks the time-reversal symmetry, and
${\cal H}_{\rm R}$ breaks the structural inversion symmetry. The last term
${\cal H}_{\rm imp}$ describes the scattering by both spin-independent and
quenched spin-dependent impurities. The latter part of ${\cal H}_{\rm imp}$
contributes to the Gilbert damping and the nonadiabatic spin
torque~\cite{Kohno:2006,Duine:2007}.

Here, we focus on effects of ${\cal H}_{\rm R}$ on the equation of
motion for the magnetization up to order $\alpha_{\rm R}$.
These effects include the DM interaction and the spin-orbit
torque. We neglect effects of order $\alpha_{\rm R}^2$ such as
interface-induced magnetic anisotropy, contributions to Gilbert
damping~\cite{Garate:2009a,Kim:2012b}, and to the nonadiabaticity
parameter~\cite{Garate:2009b}. We introduce the unitary
transformation~\cite{Aleiner:2011,Valin-Rodriguez:2011}
\begin{equation}
{\cal U}=\exp\left[ -i k_{\rm R}{\bm \sigma}\cdot ({\bf r}\times
\vhat{z})/2 \right], \label{eq:unitary-transformation}
\end{equation}
where
\begin{equation}
k_{\rm R}=\frac{2\alpha_{\rm R}m_e}{\hbar^2}
\end{equation}
and ${\bf r}=(x,y)$.
${\cal U}$ rotates the electron spin around the $\hat{\bf r}\times \vhat{z}$
direction by the angle $k_{\rm R}r$, where $r=|{\bf r}|$. We also introduce
the ${\bf r}$-dependent $3\times3$ matrix ${\cal R}$, which achieves the same
rotation of a classical vector such as $\hat{\bf m}$. Upon the unitary
transformation, one finds (Supplementary Material \cite{Supplement})
\begin{equation}
{\cal U}^\dagger {\cal H}{\cal U}
 ={\cal H}_{\rm kin}+J{\bm \sigma}\cdot\hat{\bf m}'+{\cal
H}'_{\rm imp} +{\cal
O}(\alpha_{\rm R}^2), \label{eq:2D-unitary-transformation}
\end{equation}
where
\begin{equation}
 \hat{\bf m}'  =  {\cal
R}^{-1}\hat{\bf m} \label{eq:transformed-magnetization}
\end{equation}
and
${\cal H}'_{\rm imp} = {\cal U}^\dagger {\cal H}_{\rm imp} {\cal U}$.
We ignore the last term in
Eq.~(\ref{eq:2D-unitary-transformation}) as higher order. ${\cal
H}'_{\rm imp}$ is not identical to ${\cal H}_{\rm imp}$ but they
share the same impurity expectation values up to ${\cal
O}(\alpha_{\rm R})$, which implies that ${\cal H}_{\rm R}$ has no
effect to linear order on the Gilbert damping coefficient or the
nonadiabaticity coefficient~\cite{Kohno:2006,Duine:2007}. Thus up
to ${\cal O}(\alpha_{\rm R})$, ${\cal H}'_{\rm imp}$ may be
identified with ${\cal H}_{\rm imp}$. Then the unitary
transformation from ${\cal H}$ to ${\cal U}^\dagger {\cal H}{\cal
U}$ has eliminated ${\cal H}_{\rm R}$ at the expense of replacing
$\hat{\bf m}$ by $\hat{\bf m}'$.

With this replacement, we compute the energy of
the filled Fermi sea as a function of $\hat{\bf m}$. Without
${\cal H}_{\rm R}$, the energy can depend on $\hat{\bf m}$
only through spatial derivatives $\partial_u \hat{\bf m}$ ($u=x,y$)
since the energy cannot depend on the direction of $\hat{\bf m}$ when
$\hat{\bf m}$ is homogeneous. For $\hat{\bf m}$  smoothly varying
over length scales  longer than the Fermi wavelength, the energy
density $\varepsilon$ may be expressed as the micromagnetic
exchange interaction density $\varepsilon = A \left(\partial_x
\hat{\bf m}\cdot\partial_x \hat{\bf m} + \partial_y \hat{\bf
m}\cdot \partial_y \hat{\bf m}\right)$,
where $A$ is the interfacial exchange stiffness coefficient.
Equation~(\ref{eq:2D-unitary-transformation}) implies that in the
presence of ${\cal H}_{\rm R}$, $\varepsilon$
can be obtained simply by replacing $\partial_u \hat{\bf m}$ with
$\partial_u \hat{\bf m}'$; $\varepsilon = A \left(\partial_x
\hat{\bf m}'\cdot \partial_x \hat{\bf m}' + \partial_y \hat{\bf
m}'\cdot \partial_y \hat{\bf
  m}'\right)$.
One then uses the relation (Supplementary Information)
\begin{equation}
\partial_u\hat{\bf m}'=\partial_u ({\cal R}^{-1}\hat{\bf m})
={\cal R}^{-1}\tilde{\partial}_u \hat{\bf m},
\label{eq:mprime-to-m}
\end{equation}
where the chiral derivative $\tilde{\partial}_{u}$ is defined by
\begin{equation}
\tilde{\partial}_{u}\hat{\bf m} =\partial_{u}\hat{\bf m} +k_{\rm
R} (\hat{\bf z}\times\hat{\bf u})\times \hat{\bf m}.
\label{eq:tilde-derivative-u}
\end{equation}
Here $\hat{\bf u}$ is the unit vector along the direction $u$. The
second term in Eq.~(\ref{eq:tilde-derivative-u}) arises from the
derivative operator acting on the ${\bf r}$-dependent ${\cal
R}^{-1}$.
$\varepsilon$ in the presence of the interfacial spin-orbit coupling
then becomes
\begin{eqnarray}
\varepsilon \!\! & = & \!\! A \left(\partial_x \hat{\bf m}\cdot
\partial_x \hat{\bf m}
+ \partial_y \hat{\bf m}\cdot \partial_y \hat{\bf
  m}\right) \label{eq:exchange-stiffness-R3} \\
&& \!\! +D\left[\hat{\bf y}\cdot (\hat{\bf m}\times \partial_x
\hat{\bf m}) - \hat{\bf x}\cdot (\hat{\bf m}\times \partial_y
\hat{\bf m})\right] + {\cal O}(\alpha_{\rm R}^2), \nonumber
\end{eqnarray}
with
\begin{equation}
D=2k_{\rm R}A. \label{eq:DM-coefficient}
\end{equation}
Note that the second term in Eq.~(\ref{eq:exchange-stiffness-R3})
is nothing but the interfacial DM interaction responsible for
chiral magnetic order addressed
recently~\cite{Thiaville:2012,Emori:2013,Ryu:2013}.
A few remarks are in order. First, this derivation shows that the
DM interaction is intimately related to the usual micromagnetic
exchange interaction that exists even in the absence of
interfacial spin-orbit coupling. This is the first example of the
one-to-one correspondence and illustrates how the interfacial
spin-orbit coupling generates a term in linear order from each
term present in the absence of the spin-orbit coupling.
Second, this mechanism for the DM interaction in an itinerant
ferromagnet is similar to that of the
Ruderman-Kittel-Kasuya-Yosida interaction in nonmagnetic systems
acquiring the DM-like character~\cite{Imamura:2004,Lounis:2012} when
conduction electrons are subject to interfacial spin-orbit coupling.
%

Next, we demonstrate the correlation between the DM interaction and
the spin-orbit torque. Although the spin-orbit torque has already been derived from
Eq.~(\ref{eq:RSOC-Hamiltonian}) in previous
studies~\cite{Obata:2008,Matos-Abiague:2009,Wang:2012,Kim:2012a,Pesin:2012,vanderBijl:2012},
we present below a derivation of the spin-orbit torque that
shows the relationship between it and the DM interaction.
Without ${\cal H}_{\rm R}$, it is well known~\cite{Ralph:2008}
that the total spin torque ${\bf T}_{\rm st}$ induced by an
in-plane current density ${\bf j}$ consists of the following two
components,
\begin{equation}
{\bf T}_{\rm st} = v_{\rm s}(\hat{\bf j}\cdot\bm\nabla) \hat{\bf
m}
 -\beta v_{\rm s} \hat{\bf m}\times
(\hat{\bf j}\cdot\bm\nabla) \hat{\bf
  m}, \label{eq:adia-nonadia}
\end{equation}
where the first and the second components are the
adiabatic~\cite{Tatara:2004} and
nonadiabatic~\cite{Zhang:2004,Thiaville:2005} spin toques, respectively. Here
$\hat{\bf j}={\bf j}/j$, $j=|{\bf j}|$, $\beta$ is the nonadiabaticity
parameter~\cite{Zhang:2004,Thiaville:2005}, and the spin velocity $v_{\rm s}=
P j g\mu_{\rm B} / (2 e M_{\rm s})$, where $P$ is the polarization of the
current, $g$ is the Land\'{e} $g$ factor, $\mu_{\rm B}$ is the Bohr magneton,
$M_{\rm s}$ is the saturation magnetization, and $-e$ ($<0$) is the electron
charge.
In the presence of ${\cal H}_{\rm R}$,
Eqs.~(\ref{eq:2D-unitary-transformation}) and
(\ref{eq:mprime-to-m}) imply that ${\bf T}_{\rm st}$ changes to
\begin{equation}
{\bf T}_{\rm st}  = v_{\rm s}(\hat{\bf j}\cdot\tilde{\bm\nabla})
\hat{\bf m} -\beta v_{\rm s} \hat{\bf m}\times (\hat{\bf
j}\cdot\tilde{\bm\nabla}) \hat{\bf
  m},
\label{eq:adia-nonadia-R1}
\end{equation}
where $\tilde{\bf \nabla}=(\tilde{\partial}_x,\tilde{\partial}_y)$.
One then obtains from Eq.~(\ref{eq:tilde-derivative-u})
\begin{eqnarray}
{\bf T}_{\rm st} \!\!\! &=& \!\! v_{\rm s}(\hat{\bf
j}\cdot\bm\nabla) \hat{\bf m}
-\beta v_{\rm s} \hat{\bf m}\times (\hat{\bf j}\cdot\bm\nabla)
\hat{\bf
  m} \label{eq:total-torque-Rashba} \\
&+ & \!\! \tau_{\rm f} v_{\rm s}  \hat{\bf m} \! \times \!
(\hat{\bf j}\times\hat{\bf z})
\!\! - \!\! \tau_{\rm d} v_{\rm s} \hat{\bf m}\! \times \!
[\hat{\bf m}\times(\hat{\bf j}\times\hat{\bf z})]. \nonumber
\end{eqnarray}
The two terms in the second line are the two components of the spin-orbit
torque. The first (second) component in the second line is called the
fieldlike (dampinglike) spin-orbit torque and arises from the adiabatic
(nonadiabatic) torque in the first line. This is the second example of the
one-to-one correspondence. The chiral derivative fixes the coefficients of
the two spin-orbit torque components to
\begin{equation}
\tau_{\rm f}=k_{\rm R}, \ \tau_{\rm d}=\beta k_{\rm R}.
\label{eq:SOT-magnitudes}
\end{equation}
When combined with Eq.~(\ref{eq:DM-coefficient}), one finds
\begin{equation}
\tau_{\rm f}=D/2A, \, \tau_{\rm d}=\beta D/2A.
\label{eq:D-tau-correlation}
\end{equation}
This correlation between the DM coefficient $D$ and the spin-orbit torque
coefficients $\tau_{\rm f}$ and $\tau_{\rm d}$ is a key result of this
work.

A recent experiment~\cite{Emori:2013} examined current-driven
domain wall motion in the systems Pt/CoFe/MgO and Ta/CoFe/MgO and
concluded that domain wall motion against (along) the electron flow in
the former (latter) system is due to the product $D\tau_{\rm d}P$
being positive (negative).
According to Eqs.~(\ref{eq:DM-coefficient}) and
(\ref{eq:SOT-magnitudes}), $D\tau_{\rm d}P=2\beta P A k_{\rm R}^2$
should be of the same sign as $\beta P$ regardless of $k_{\rm R}$
since $A$ is positive by definition.
Thus explaining the experimental results for Ta/CoFe/MgO within the
interfacial spin-orbit coupling theory requires $\beta P$ to be negative.
Whereas $\beta P$ can be negative, in most models and parameter ranges it is
positive. We tentatively conclude that $\tau_{\rm d}$ in
Ta/CoFe/MgO~\cite{Emori:2013} has a different origin, the spin Hall effect
being a plausible mechanism as argued in Ref.~\cite{Emori:2013}.
For Pt/CoFe/MgO, on the other hand, the reported sign is
consistent with the sign determined from
Eqs.~(\ref{eq:DM-coefficient}) and (\ref{eq:SOT-magnitudes}) if
$\beta P>0$. The Pt-based structure in Ref.~\cite{Ryu:2013} also
gave the same sign as Ref.~\cite{Emori:2013}.
To investigate the origin of the spin-orbit torque in Pt/CoFe/MgO, we attempt
a semiquantitative analysis. For the suggested values $D=0.5$~mJ/m$^2$,
$A=10^{-11}$~J/m in Ref.~\cite{Emori:2013}, Eq.~(\ref{eq:DM-coefficient})
predicts $k_{\rm R}=2.5\times 10^8$~m$^{-1}$. For $P=0.5$, $\beta=0.4$,
$M_{\rm s}=3\times 10^5$~Am$^{-1}$, which are again from
Ref.~\cite{Emori:2013}, Eq.~(\ref{eq:SOT-magnitudes}) predicts the effective
transverse field $-(\tau_{\rm f}v_{\rm s}/\gamma)\hat{\bf j}\times \hat{\bf
z}$ of the fieldlike spin-orbit torque and the effective longitudinal field
$(\tau_{\rm d}v_{\rm s}/\gamma)(\hat{\bf m}\times(\hat{\bf m}\times \hat{\bf
z}))$ of the dampinglike spin-orbit torque to have the magnitudes 1.3 mT and
0.52 mT, respectively, for $j=10^{11}$~A/m$^2$. Here $\gamma$ is the
gyromagnetic ratio. The former value is in reasonable agreement with the
measured value 2 mT considering uncertainty in the parameter values quoted
above, whereas the latter value is about an order of magnitude smaller than
the measured value 5 mT in Ref.~\cite{Emori:2013}. We thus conclude that the
fieldlike spin-orbit torque of Pt/CoFe/MgO in Ref.~\cite{Emori:2013} is
probably due to the interfacial spin-orbit coupling whereas the dampinglike
spin-orbit torque is probably due to a different mechanism such as the bulk
spin Hall effect. For the fieldlike spin-orbit torque of Pt/CoFe/MgO, the
relative sign of $\tau_{\rm f}$ with respect to $D$ is also consistent with
the prediction of the interfacial spin-orbit coupling if $P$ is positive.

\begin{figure}
\includegraphics[width=8.6cm]{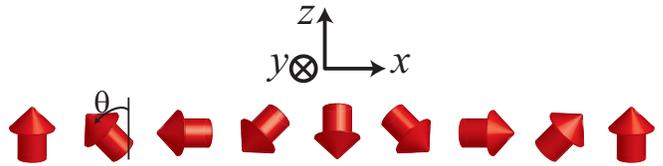}
\caption{(color online) Chiral precession of magnetization
$\hat{\bf m}$. Chiral precession profile of $\hat{\bf m}$ with
$\tilde{\partial}_x \hat{\bf m}=0$ forms a left-handed (for
$k_{\rm R}>0$) cycloidal spiral. This profile is identical to the
spin precession profile of conduction electrons moving in the $+x$ or
$-x$ direction in nonmagnetic systems with ${\cal H}_{\rm
R}$~\cite{Datta:1990}.
 }
\label{fig:1d-curved-spin-space}
\end{figure}

These two examples illustrate the idea that all linear effects of the
interfacial spin-orbit coupling can be captured through the chiral derivative
$\tilde{\partial}_u \hat{\bf m}$. To gain insight into its physical meaning,
it is illustrative to take $u=x$ and examine the solution of
$\tilde{\partial}_x \hat{\bf m}=0$, which forms a left-handed (for $k_{\rm
R}>0$) cycloidal spiral (Fig.~\ref{fig:1d-curved-spin-space}), where
$\hat{\bf m}$ precesses around the $-(\hat{\bf z}\times \hat{\bf x})$ axis
[$-(\hat{\bf z}\times \hat{\bf y})$ axis if $u=y$] as $x$ increases with the
precession rate $d\theta/dx=k_{\rm R}$. This chiral precession gives the
name, chiral derivative.
Note that this precession is identical to the conduction electron spin
precession caused by ${\cal H}_{\rm R}$ in nonmagnetic
systems~\cite{Datta:1990}.
Moreover when $\tilde{\partial}_x \hat{\bf m}=0$, ${\cal H}_{\rm
exc}$ also causes the same conduction electron spin precession as
${\cal H}_{\rm R}$ does.
%
%
Thus effects of ${\cal H}_{\rm R}$ and ${\cal H}_{\rm exc}$ become
harmonious and
the one-dimensional ``half" $p_x^2/2m_e-(\alpha_{\rm
R}/\hbar)\sigma_y p_x + J{\bm \sigma}\cdot \hat{\bf m}$ of the 2D
Hamiltonian~(\ref{eq:RSOC-Hamiltonian}) gets minimized when
$\tilde{\partial}_x \hat{\bf m}=0$. Interestingly, the sum of the
exchange energy and the DM interaction, namely
$A\partial_x\hat{\bf m}\cdot
\partial_x \hat{\bf m}+D(\hat{\bf z}\times \hat{\bf
x})\cdot(\hat{\bf m}\times\partial_x\hat{\bf m})$, also gets
minimized when $\tilde{\partial}_x \hat{\bf m}=0$. This is not a
coincidence as this sum by definition should agree with the energy
landscape of the Hamiltonian, which forces the value $D$ in
Eq.~(\ref{eq:DM-coefficient}).

One consequence of deriving the spin-orbit torque using the chiral derivative
is that such a derivation shows that the spin-orbit torque is chiral when
combined with the conventional spin torque just as the DM interaction is
chiral when combined with the micromagnetic exchange interaction. For
example, when ${\bf j}$ is along the $x$ direction, the total torque ${\bf
T}_{\rm st}$ in Eq.~(\ref{eq:adia-nonadia-R1}) vanishes even for finite $j$
if $\tilde{\partial}_x \hat{\bf m}=0$.
As a side remark, the first and second terms in
Eq.~(\ref{eq:adia-nonadia-R1}) are nothing but current-dependent corrections
to the torques due to the total equilibrium energy density in
Eq.~(\ref{eq:exchange-stiffness-R3}) and the Gilbert damping, respectively.
This identification is a straightforward generalization of a previously
reported counterpart; when ${\cal H}_{\rm R}$ is absent, the adiabatic and
nonadiabatic spin torques in Eq.~(\ref{eq:adia-nonadia}) are the
current-dependent corrections to the torques due to the micromagnetic
exchange interaction~\cite{Xiao:2006} and the Gilbert
damping~\cite{Garate:2009a}.

The anomalously fast current-driven domain wall motion demonstrated in
Ref.~\cite{Ryu:2013} raises the possibility that chirally ordered magnetic
structures~\cite{Bode:2007,Heinze:2011} such as topological Skyrmion lattices
may be very efficiently controlled electrically. Such motion would be similar
to the highly efficient electrically driven dynamics of a Skyrmion lattice in
a system with bulk spin-orbit coupling such as the B20
structure~\cite{Jonietz:2010}.
Flexible deformation of the Skyrmion lattice is proposed~\cite{Iwasaki:2013}
as an important contribution to the high efficiency of current-driven
dynamics in B20 structures. We expect Skyrmion lattices in magnetic bilayers
to behave similarly because both systems are similarly frustrated. The chiral
derivative is noncommutative, $\tilde{\partial}_x \tilde{\partial}_y \hat{\bf
m}\neq \tilde{\partial}_y \tilde{\partial}_x \hat{\bf m}$, so the energy
landscape of the lattice structure is necessarily frustrated leading to the
existence of many metastable structures with low excitation energies.
%

In a Skyrmion lattice, another linear effect of the interfacial spin-orbit
coupling becomes important.
Consider a Skyrmion lattice without interfacial spin-orbit coupling. The
spatial variation of $\hat{\bf m}$ introduces a real space Berry
phase~\cite{Neubauer:2009}, which can affect the electron transport through a
Skyrmion lattice. It produces a fictitious magnetic field~\cite{Volovik:1987}
${\bf B}^\pm=\mp (h/e)\hat{\bf z}b$, where $b=(\partial_x \hat{\bf m}\times
\partial_y \hat{\bf m})\cdot \hat{\bf m}/4\pi$ is nothing but the Skyrmion
number density~\cite{Neubauer:2009}.
Here the upper and lower signs apply to majority (spin antiparallel to
$\hat{\bf m}$) and minority (spin parallel to $\hat{\bf m}$) electrons, and
thus this field is spin dependent.
An experiment~\cite{Heinze:2011} on Fe/Ir bilayer reported the Skyrmion
spacing of 1 nm.  For a Skyrmion density of (1 nm)$^{-2}$, ${\bf B}^\pm$
becomes of the order of $10^4$ T, which can significantly affect electron
transport.

In the presence of interfacial spin-orbit coupling, the
Berry-phase-derived field becomes chiral.
Following the same procedure as above, one finds that ${\bf
B}^\pm$ is now given by $\mp (h/e)\hat{\bf z}\tilde{b}$, where
$\tilde{b}=(\tilde{\partial}_x \hat{\bf m}\times
\tilde{\partial}_y \hat{\bf m})\cdot \hat{\bf m}/4\pi=b+b_{\rm
R}+{\cal O}(\alpha_{\rm R}^2)$, where
\begin{equation}
b_{\rm R}= k_{\rm R}{\bm \nabla}\cdot\hat{\bf m}/4\pi.
\label{eq:spin-dependent-B-field-RSOC}
\end{equation}
We estimate the magnitude of $b_{\rm R}$ for the Mn/W
bilayer~\cite{Bode:2007}, for which left-handed cycloidal spiral with period
12 nm is reported. From the estimated value $D=23.8/(2\pi)$ nm meV per Mn
atom and $A=94.2/(2\pi)^2$ nm$^2$ meV per Mn atom, we find $k_{\rm R}=0.794$
nm$^{-1}$ from Eq.~(\ref{eq:DM-coefficient}), and $(h/e)b_{\rm R}$ becomes
about 140 T. Thus for the left-handed cycloidal spiral, for which the
Skyrmion density $b=0$, the effective magnetic field is governed by this
interfacial spin-orbit coupling contribution.
%

For completeness, we also discuss briefly the interfacial spin-orbit coupling
contribution to the fictitious electric field ${\bf E}^\pm$, which is spin
dependent and arises when $\hat{\bf m}$ varies in time.
Without ${\cal H}_{\rm R}$, it is known that ${\bf
E}^\pm=\pm(h/4\pi e)({\bf e}^{\rm adia}+{\bf e}^{\rm non})$, where
the so-called adiabatic
contribution~\cite{Volovik:1987,Barnes:2007,Yang:2010} is given by
$({\bf e}^{\rm adia})_u  = (\partial_t \hat{\bf m}\times
\partial_u \hat{\bf m})\cdot \hat{\bf m}$ and the nonadiabatic
contribution~\cite{Duine:2008,Tserkovnyak:2008} is given by $({\bf e}^{\rm
non})_u = \beta (\partial_u \hat{\bf m}\cdot
\partial_t \hat{\bf m})$.
In the presence of ${\cal H}_{\rm R}$, corrections arise. Recently
some of us~\cite{Kim:2012b} reported a correction term ${\bf
e}^{\rm adia}_{\rm R}$, and Ref.~\cite{Tatara:2013} reported
another correction term ${\bf e}^{\rm non}_{\rm R}$, which are
given by
\begin{eqnarray}
({\bf e}^{\rm adia}_{\rm R})_u & = & -k_{\rm R} (\vhat{z}\times
\vhat{u})\cdot
\partial_t \hat{\bf m},
\label{eq:SD-E-field-field-like} \\
({\bf e}^{\rm non}_{\rm R})_u & = &  \beta k_{\rm R}
(\vhat{z}\times \vhat{u})\cdot (\hat{\bf m} \times \partial_t
\hat{\bf m}). \label{eq:SD-E-field-damping-like}
\end{eqnarray}
Here we point out that the previously reported corrections can be
derived almost trivially using the chiral derivative since
$({\bf e}^{\rm adia}+{\bf e}^{\rm adia}_{\rm R})_u  = (\partial_t
\hat{\bf m}\times \tilde{\partial}_u \hat{\bf m})\cdot \hat{\bf
m}$ and $({\bf e}^{\rm non}+{\bf e}^{\rm non}_{\rm R})_u  =  \beta
(\tilde{\partial}_u \hat{\bf m}\cdot
\partial_t \hat{\bf m})$. This derivation
also reveals the chiral nature of ${\bf e}^{\rm adia}_{\rm R}$ and
${\bf e}^{\rm non}_{\rm R}$.
For the drift motion of chiral magnetic structures at 100 m/s, the
parameter values of the Mn/W bilayer~\cite{Bode:2007} lead to the
estimation that both $(h/4\pi e)({\bf e}^{\rm adia})$ and $(h/4\pi
e)({\bf e}^{\rm adia}_{\rm R})$ are of the order of $10^4$ V/m,
which should be easily detectable.

So far we focused on magnetic bilayers. But
these results should also be relevant for the high-mobility 2D electron
gas formed at the interface between two different insulating oxide
materials. One example is the LaAlO$_3$/SrTiO$_3$
interface~\cite{Ohtomo:2004}, which has broken
structural inversion symmetry~\cite{Caviglia:2010} and becomes
magnetic~\cite{Brinkman:2007} under proper conditions.

Last, we briefly discuss how features of real systems might affect our
conclusions. Two differences in realistic band structures, are that the
energy-momentum dispersion is not parabolic and that there are multiple
energy bands~\cite{Park:2013}. Another difference is that magnetic bilayers
are not strictly 2D systems, unlike systems such as LaAlO$_3$/SrTiO$_3$.
To test the effects of more realistic band structures, in the Supplementary
Material \cite{Supplement}, we examine a tight-binding version of ${\cal H}$,
which generates nonparabolic energy bands, and find that the
relation~(\ref{eq:DM-coefficient}) remains valid despite the nonparabolic
dispersion.
The two dimensionality is tested in a recent publication by some of
us~\cite{Haney:2013a}.  There, we perform a {\it three}-dimensional Boltzmann
calculation to address the interfacial spin-orbit coupling effect on the
spin-orbit torque and obtain results, which are in qualitative agreement with
those of the 2D Rashba model.
On the basis of these observations, we expect that predictions of the simple
Rashba model will survive at least qualitatively even in realistic situations
and thus can serve as a good reference point for more quantitative future
analysis.

To conclude, we examined effects of interfacial spin-orbit coupling using the
Rashba model.  We found that all linear effects of the interfacial spin-orbit
coupling can be derived by replacing spatial derivatives with chiral
derivatives.  This allows these effects to be understood in terms of chiral
generalizations of effects in the absence of spin-orbit coupling. One
important consequence is a relationship between the DM interaction and the
spin-orbit torque, such that measuring one should give a strong indication of
the other.

\begin{acknowledgements}
H.W. L. wishes to thank Ki-Seok Kim, F. Freimuth, S. Bl\"ugel, T. Silva, and
P. Haney for critical comments and useful discussions. Financial support was
provided by NRF Grants No. 2013R1A2A2A05006237 and No. 2011- 0030046 (H.W.
L.) and No. 2013R1A2A2A01013188 (K. J. L.).
\end{acknowledgements}

\end{document}